\documentclass[10pt]{article} 
\def\fl{} 
\def \D {\hbox{d}}

\begin{document}

\title
{Painlev\'e structure of a multi-ion electrodiffusion system\footnote
{Preprint S2007/023. To appear, Journal of Physics A. nlin.SI/yymmnnn}}

\author{R Conte$^1$, C Rogers$^{2,4}$ and W K Schief$^{3,4}$
{}\\
$^1$Service de physique de l'\'etat condens\'e (URA 2464),
\\ CEA--Saclay, F--91191 Gif-sur-Yvette Cedex, France
{}\\ \\
$^2$Department of Applied Mathematics,
\\ Hong Kong Polytechnic University, Kowloon, Hong Kong
{}\\ \\
$^3$Institut f\"ur Mathematik, Technische Universit\"at Berlin,
\\ Stra{\ss}e des 17.\ Juni 136, D--10623 Berlin, Germany
{}\\ \\
$^4$Australian Research Council Centre of Excellence for Mathematics
\\ and Statistics of Complex Systems, School of Mathematics and Statistics,
\\ The University of New South Wales, Sydney, NSW 2052, Australia
\\ \\
Robert.Conte@cea.fr, colinr@maths.unsw.edu.au, schief@math.tu-berlin.de
}

\maketitle

\begin{abstract}
A nonlinear coupled system descriptive of multi-ion electrodiffusion
is investigated and all parameters for which the system admits a
single-valued general solution are isolated.
This is achieved \textit{via} a method initiated by Painlev\'e
with the application of a test due to Kowalevski and Gambier.
The solutions can be obtained explicitly in terms of Painlev\'e
transcendents or elliptic functions.
\end{abstract}

PACS: 02.20.Qs, 11.10.Lm
\baselineskip=12truept 

\section{Introduction}

The theory of electrodiffusion has its origin in the liquid-junction
theory of Nernst and Planck \cite{1}.
It seeks to provide a macroscopic description of the transport
of charged particles through material barriers.
Applications abound,
in particular, in the modelling of biological membranes
\cite{2}-\cite{7}
and in electrochemistry \cite{8}.
Schl\"ogl \cite{9}
observed that it is convenient to partition the ions
into $m$ classes characterized by the same electric charge
$q_j=q_0 \nu_j$,
where $q_0$ is the unit of charge
and $\nu_j$ is a nonzero integral signed valency.
The $m$-ion electrodiffusion model in steady r\'egimes then
reduces to the following system of $m+1$ coupled first order
ODEs \cite{10}:
\begin{eqnarray}
& &
\left\lbrace
\begin{array}{ll}
\displaystyle{
\frac{\D n_i}{\D x} = \nu_i n_i p - c_i,\quad \nu_i n_i \not=0,\quad i=1,\cdots,m,
}
\\[2ex]
\displaystyle{
\frac{\D p}{\D x} = \sum_{i=1}^{m} \nu_i n_i,
}
\end{array}
\right.
\label{eqSystem_ni_p}
\end{eqnarray}
where
$x$ is the coordinate normal to the planar boundaries,
$p$ is the electric field and
$n_j$ is the number of ions with the same charge $q_j=q_0 \nu_j$.
These variables are subject to  the constraint \cite{9}
\begin{eqnarray}
& &
(\nu_i-\nu_j) \nu_i n_i \not=0,\quad  i \ne j. \
\label{eqRestriction}
\end{eqnarray}

The system (\ref{eqSystem_ni_p})
admits the first integral
\begin{eqnarray}
& &
K = \frac{p^2}{2} - Cx - \sum_{i=1}^{m} n_i,\
\label{eqSystem_ni_pFirstIntegral}
\end{eqnarray}
where $K$  may be set to zero without loss of generality
whenever $ C = \sum\limits_{i=1}^m c_i \ne 0$.
Note that certain  boundary value problems
indeed require that $K$ be zero.
Here, this constraint is not imposed and
it is demonstrated in an algorithmic manner that the
electrodiffusion model for
$m\leq 4$ admits solutions given in terms of Painlev\'e transcendents and
elliptic functions.
The Painlev\'e reduction obtained in \cite{Bass}
is retrieved as a particular case.

\section{Single-valued solutions. A classical procedure}
\label{sectionClassical}

For nonlinear systems such as (\ref{eqSystem_ni_p}),
dependent on certain parameters $\nu_i,c_i$,
there is a classical method \cite{PaiBSMF}
for determining all those parameters for which the general solution
of the nonlinear system is single-valued around its movable singularities.
An alternative approach due to Kowalevski \cite{Kowa1889}
and Gambier \cite{GambierThese}
is also available,
and all the details which are sometimes missing in these classical authors
can be found in \cite{BureauMI}
or in summer school lecture notes \cite{Cargese1996Conte}.

The main steps, to be performed in the sequel, are as follows:
\begin{enumerate}
\item
A scaled version of the nonlinear system (\ref{eqSystem_ni_p})
is introduced for new fields $N_i(X),P(X)$,
\textit{via} the transformation
(this is a particular $\alpha$-transformation \cite{PaiBSMF}),
\begin{eqnarray}
& &
x=\varepsilon X,\quad
p=\varepsilon^{-1} P,\quad
n_i=N_i.
\label{eq_alpha_transformation}
\end{eqnarray}
Thus the limit $\varepsilon \to 0$ generates
\begin{eqnarray}
& &
\left\lbrace
\begin{array}{ll}
\displaystyle{
\frac{\D N_i}{\D X} = \nu_i N_i P,\quad \nu_i \not=0,\quad i=1,\cdots,m,
}
\\[2ex]
\displaystyle{
\frac{\D P}{\D X} = \sum_{i=1}^{m} \nu_i N_i,
}
\\[2ex]
\displaystyle{
0 = \frac{P^2}{2} - \sum_{i=1}^{m} N_i.
}
\end{array}
\right.
\label{eqSystem_ni_p-simplified}
\end{eqnarray}
This canonical system results from (\ref{eqSystem_ni_p})
and (\ref{eqSystem_ni_pFirstIntegral})
by setting $c_i=0, K=0$
and is termed the \textit{simplified system},
as opposed to the original \textit{complete system} (\ref{eqSystem_ni_p}).
If the complete system has a single-valued general solution,
it is necessary that the simplified system has
a single-valued general solution.

\item
Necessary conditions on the parameters $\nu_i$
for the general solution of the simplified system to be single-valued
are determined.
These conditions are shown to be sufficient for the problem at hand.

\item
The necessary conditions on the parameters $\nu_i,c_i$
for the general solution of the complete system to be single-valued
are obtained and the corresponding PII and other reductions listed.

\end{enumerate}

\section{Constraints on the charges $\nu_i$}
\label{sectionDiophante}

The transformation (\ref{eq_alpha_transformation})
implies that $P(X)$ admits movable simple poles,
\begin{eqnarray}
& &
P=\chi^{-1}\left[P_0+\mathcal{O}(\chi)\right],\quad \chi=X-X_0,\quad
P_0 \not=0.
\label{eqleadingP}
\end{eqnarray}
A straightforward integration of (\ref{eqSystem_ni_p-simplified})${}_1$
reveals the leading order behaviour
\begin{eqnarray}
& &
\left\lbrace
\begin{array}{ll}
\displaystyle{
P=\chi^{-1}\left[P_0+\mathcal{O}(\chi)\right],\quad 
}
\\[2ex]
\displaystyle{
N_i=\chi^{\nu_i P_0}\left[A_i+\mathcal{O}(\chi)\right],
}
\\[2ex]
\displaystyle{
0 = \frac{P_0^2}{2} \chi^{-2} \left[1+\mathcal{O}(\chi)\right]
- \sum_{i=1}^{m} \chi^{\nu_i P_0}\left[A_i+\mathcal{O}(\chi)\right].
}
\end{array}
\right.
\label{eqsimplifiedSolutionLocal}
\end{eqnarray}
Here,
the $A_i$ are undetermined constants.
Since the $\nu_i$ are assumed distinct (see (\ref{eqRestriction})),
the term in $\chi^{-2}$ in (\ref{eqsimplifiedSolutionLocal})${}_3$
can only be matched by exactly one of the terms involving
$\chi^{\nu_i P_0}$.
Denoting $j$ this matching index,
this yields the desired value of $P_0$,
\begin{eqnarray}
& &
P_0=-\frac{2}{\nu_j},\quad
A_j= \frac{2}{\nu_j^2}.
\label{eq_leading_coeffj}
\end{eqnarray}
In view of the possible vanishing of the remaining constants $A_i$,
the necessary condition that each $N_i(X)$ be single-valued near $X=X_0$
is not easy to enforce and hence
we proceed directly with the other conditions.

An additional necessary condition
is that the linearized version of the simplified system
near the local behaviour (\ref{eqsimplifiedSolutionLocal})
also has the property that
its general solution is single-valued near $X=X_0$.
Since this linearized system is of Fuchsian type
\cite[Chap.~XVI]{Ince} near $X=X_0$,
it is necessary that all its Fuchs indices be integer.
This condition is here readily enforced with knowledge of $P_0$.
Thus, let us introduce the function $W$ by
\begin{eqnarray}
& &
P=-2 \frac{W'}{\nu_j W}.
\label{eqspt}
\end{eqnarray}
By construction, $W$ has a simple zero near $X=X_0$
(it can be chosen such that
\hfill\break\noindent
$\lim_{X \to X_0} W/(X-X_0) = 1$).
The integration of the simplified system then reduces
to the integration of a single ODE for $W(X)$, namely
\begin{eqnarray}
& &
{W'}^2
- \frac{\nu_j^2}{2} \sum_{i=1}^{m} k_i \ W^{2 -2 \nu_i / \nu_j}=0,
\label{eqBriotBouquetW}
\end{eqnarray}
where
\begin{eqnarray}
& &
N_i=k_i W^{-2 \nu_i / \nu_j},
\label{eqNiof W}
\end{eqnarray}
and the $k_i$ are arbitrary constants.
Since $\lim_{X \to X_0} W/(X-X_0) = 1$,
Laurent series in $X-X_0$ are also Laurent series in $W$.
The values of the Fuchs indices are directly read from
(\ref{eqBriotBouquetW}),
and are given by
\begin{eqnarray}
& &
\hbox{Fuchs indices = }
-1, 2-2\frac{\nu_i}{\nu_j},\quad i=1,\dots,m.
\label{eqpindices}
\end{eqnarray}
Finally,
since $j$ can be arbitrarily chosen,
the diophantine condition to be solved is
\begin{eqnarray}
& &
\forall j\
\forall i:\
-2 \frac{\nu_i}{\nu_j}= \hbox{ integer}.
\label{eqDiophante}
\end{eqnarray}
The latter condition is restrictive
and, using the shorthand notation $\nu_1:\nu_2:\nu_3 = a:b:c$ to indicate
\begin{eqnarray}
& &
\frac{\nu_1}{a}=\frac{\nu_2}{b}=\frac{\nu_3}{c}=\hbox{arbitrary},
\end{eqnarray}
its only distinct solutions are
\begin{eqnarray}
& &
\left\lbrace
\begin{array}{ll}
\displaystyle{
m=1:\quad \hbox{no restriction},
}
\\[1ex]
\displaystyle{
m=2:\quad \nu_1:\nu_2=1:-2,\quad                 1:-1,\quad              1:2,
}
\\[1ex]
\displaystyle{
m=3:\quad \nu_1:\nu_2:\nu_3     = 1:-2:-1,\quad          1:-2:2,
}
\\[1ex]
\displaystyle{
m=4:\quad  \nu_1:\nu_2:\nu_3:\nu_4 = 1:-2:-1:2,
}
\\[1ex]
\displaystyle{
m > 4:\quad \hbox{no solution}.
}
\end{array}
\right.
\label{eqlistmnu}
\end{eqnarray}

The necessary condition (\ref{eqDiophante})
is also proven to be sufficient.
Thus,
in the first order ODE (\ref{eqBriotBouquetW}) for $W(X)$,
all the powers of $W$ are integers between $0$ and $4$
for the values (\ref{eqlistmnu})
so that this ODE
belongs to the binomial type studied by Briot and Bouquet
\cite[pages 58--59]{PaiLecons}
and its general solution $W(X)$ is single-valued.
The explicit expressions for $P$ and $N_i$
in terms of $W$ prove the single-valuedness of the general
solution of the simplified system
(\ref{eqSystem_ni_p-simplified}).
This completes the second step as set down in Section \ref{sectionClassical}.

Let us remark that the Fuchs indices (\ref{eqpindices}) comprise a subset
of $\{-2,-1,1,3,4,6\}$.

\section{Constraints on the $c_i$}
\label{sectionNolog}

The third step involves a well known test
due to Kowalevski and Gambier.
In view of the numerous possible dominant behaviours (\ref{eqNiof W})
of the $n_i$,
it proves convenient to eliminate the $m$ variables $n_i$ to obtain
the $m$-th order ODE obeyed by the electric field $p(x)$.
Thus, taking the derivative of (\ref{eqSystem_ni_p})$_2$ $m-1$ times,
one generates a van der Monde system for the $m$ variables $n_i$,
namely \cite{10}
\begin{eqnarray}
& & \left\lbrace
\begin{array}{ll}
\displaystyle{
\sum_{i=1}^m \nu_i^q n_i = a_q,\quad q=1,\cdots,m
}
\\
\displaystyle{
a_1=p',\
p a_q = a_{q-1}' + \sum_{k=1}^m \nu_k^{q-1} c_k,\quad q=2,\cdots,m.
}
\end{array}
\right.
\label{eqvanderMonde}
\end{eqnarray}
In view of the constraint (\ref{eqRestriction}),
the solution $n_i$ of this system is unique.
Thus,
the question of single-valuedness of the general solution $(n_i,p)$
of the complete system (\ref{eqSystem_ni_p}) reduces to that of $p$ alone.
The $m$-th order first-degree ODE for $p(x)$ is
defined by the determinant
\begin{eqnarray}
& &
\left|\matrix{
 1 & \dots   & 1 & p^2/2 - (C x +K) \cr
 \nu_1   & \dots   & \nu_m   & a_1   \cr
 \dots   & \dots   & \dots   & \dots \cr
 \nu_1^q & \dots   & \nu_m^q & a_q   \cr
 \dots   & \dots   & \dots   & \dots \cr
 \nu_1^m & \dots   & \nu_m^m & a_m   \cr
}\right|
=0.
\label{eqode_p_orderm}
\end{eqnarray}
For the admissible values of $m $ as set down in (\ref{eqlistmnu}),
one obtains \cite{10}:
\begin{eqnarray}
& &
\left\lbrace
\begin{array}{ll}
\displaystyle{
m=1:\quad
p'+ \nu_1 \left[-\frac{p^2}{2} + C x + K\right]=0,
}
\\[2ex]
\displaystyle{
m=2:\quad
p'' - (\nu_1 + \nu_2) p p'
 + \nu_1 \nu_2 \left[\frac{p^3}{2} -(C x + K) p\right]
 + \nu_1 c_1 + \nu_2 c_2=0,
}
\\[2ex]
\displaystyle{
m=3:\quad
p p''' - p' p'' - (\nu_1 + \nu_2 + \nu_3) p^2 p''
+ (\nu_2 \nu_3 + \nu_3 \nu_1 + \nu_1 \nu_2) p^3 p'
} \\[2ex]
 \displaystyle{\phantom{1234567}
+ \nu_1 \nu_2 \nu_3
\left[-\frac{p^5}{2} +(C x + K) p^3 \right]
} \\[2ex]
 \displaystyle{\phantom{1234567}
-\left[
 (\nu_2+\nu_3)\nu_1 c_1 + (\nu_3+\nu_1)\nu_2 c_2 + (\nu_1+\nu_2)\nu_3 c_3
\right] p^2
} \\[2ex]
 \displaystyle{\phantom{1234567}
- (c_1 \nu_1+c_2 \nu_2 + c_3 \nu_3) p'
=0,
}
\\[2ex]
\displaystyle{
m=4:\quad
p^2 p'''' - 3 p p' p''' + 3 {p'}^2 p'' - p {p''}^2
} \\[2ex]
 \displaystyle{\phantom{1234567}
+ s_1 (-p^3 p''' + p^2 p' p'')
+ s_2 p^4 p''
- s_3 p^5 p'
+\frac{s_4}{2} p^7
- s_4 (C x +K) p^5
} \\ \displaystyle{\phantom{1234567}
+ A_3 p^4
+ A_1 (3 {p'}^2 - p p'')
+ A_2 p^2 p'
=0,
}
\end{array}
\right.
\label{eqode_p_orders1234}
\end{eqnarray}
with the notation
\begin{eqnarray}
& & \left\lbrace
\begin{array}{ll}
\displaystyle{
s_1=\nu_1+\nu_2+\nu_3+\nu_4,
}
\\
\displaystyle{
s_2=\nu_1 \nu_2+\nu_1 \nu_3+\nu_1 \nu_4+\nu_2 \nu_3+\nu_2 \nu_4+\nu_3 \nu_4,
}
\\[1ex]
\displaystyle{
s_3=\nu_2 \nu_3 \nu_4+\nu_1 \nu_3 \nu_4+\nu_1 \nu_2 \nu_4+\nu_1 \nu_2 \nu_3,
}
\\[1ex]
\displaystyle{
s_4=\nu_1 \nu_2 \nu_3 \nu_4,
}
\\[1ex]
\displaystyle{
A_3=(c_2+c_3+c_4) \nu_2 \nu_3 \nu_4
   +(c_1+c_3+c_4) \nu_1 \nu_3 \nu_4
} \\[1ex]
 \displaystyle{\phantom{123}
   +(c_1+c_2+c_4) \nu_1 \nu_2 \nu_4
   +(c_1+c_2+c_3) \nu_1 \nu_2 \nu_3,
}
\\[1ex]
\displaystyle{
A_2=(c_1+c_2) \nu_1 \nu_2
   +(c_1+c_3) \nu_1 \nu_3
   +(c_1+c_4) \nu_1 \nu_4
} \\[1ex]
 \displaystyle{\phantom{123}
   +(c_2+c_3) \nu_2 \nu_3
   +(c_2+c_4) \nu_2 \nu_4
   +(c_3+c_4) \nu_3 \nu_4,
}
\\[1ex]
\displaystyle{
A_1=c_1 \nu_1+c_2 \nu_2+c_3 \nu_3+c_4 \nu_4.
}
\end{array}
\right.
\label{eqode_notation}
\end{eqnarray}

The next set of necessary conditions
arises when one investigates the existence of the $m$ Laurent series
(one for each chosen $j$)
whose first term is (see (\ref{eqleadingP}))
\begin{eqnarray}
& &
p = \sum_{k=0}^{+\infty} p_k (x-x_0)^{k-1},\quad p_0=-\frac{2}{\nu_j},\quad
j=1,\cdots,m.
\label{eqpLaurent}
\end{eqnarray}
The locations $k$ at which arbitrary constants may appear in the series
(\ref{eqpLaurent})
are identical
to the Fuchs indices listed in (\ref{eqpindices}).

Whenever $k$ reaches one of the positive Fuchs indices
in the list (\ref{eqpindices}),
to eliminate the possibility of
a movable logarithm in the expansion (\ref{eqpLaurent})
destructive of the Painlev\'e property,
conditions must be imposed on the $c_i,  \, i = 1,\cdots, m.$
The results of this classical computation are (Appendix):
\begin{eqnarray}\fl
& & \left\lbrace
\begin{array}{ll}
\displaystyle
m=1:&\quad \hbox{no additional condition},
\\
\displaystyle
m=2,\quad \nu_1:\nu_2=1:-2:&\quad c_1=0,\quad c_2=0,
\\
\displaystyle
m=2,\quad \nu_1:\nu_2=1:-1:&\quad  \hbox{no additional condition},
\\
\displaystyle
m=2,\quad \nu_1:\nu_2=1:2:&\quad   \hbox{no additional condition},
\\
\displaystyle
m=3,\quad \nu_1:\nu_2:\nu_3 = 1:-2:-1:&\quad  c_1=0,\quad c_2(2 c_2+3 c_3)=0,
\\
\displaystyle
m=3,\quad \nu_1:\nu_2:\nu_3 = 1:-2:2:&\quad c_1=0,\quad c_2=0,
\\
\displaystyle
m=4,\quad  \nu_1:\nu_2:\nu_3:\nu_4 = 1:-2:-1:2:&\quad c_1=c_2=c_3=c_4=0.
\end{array}
\right.
\label{eqPositiveLog}
\end{eqnarray}
The value $K$ of the first integral is unconstrained.

Whenever $-2$ is a Fuchs index of some family,
movable logarithms may occur as well from this index $-2$,
but the classical computation recalled in the Appendix
cannot detect them.
In order to perform this detection,
an analysis up to the perturbation order
$n=6$ has been conducted in the manner described in \cite{13}.
All such additional necessary conditions
turn out to be  identically satisfied.

In order to complete the third step,
it remains to examine whether the general solution $(n_i,p)$
of the complete system
is indeed single-valued for the cases set down in (\ref{eqPositiveLog}).
It proves convenient
to make use of one of the $m$ variables $w$
involved in the singular part transformation (see (\ref{eqspt})),
\begin{eqnarray}
& &
p=-2 \frac{w'}{\nu_j w},
\label{eqsptComplete}
\end{eqnarray}
namely the one
where  $\vert \nu_j \vert$ is the greatest of the $m$ charges.
To each vanishing $c_i$, if any,
there corresponds a first integral
\begin{eqnarray}
& &
k_i =  n_i w^{2 \nu_i / \nu_j} \not=0,
\label{eqfirstki}
\end{eqnarray}
where the  exponent $2\nu_i/\nu_j$ is an integer.
For each  pair  $(c_i,c_k)$ with $c_i = c_k =0$
one can eliminate $w$ between the two associated first integrals
to obtain a first integral rational in $(n_i,n_k)$, namely
\begin{eqnarray}
& &
n_i^{2 \nu_j/\nu_i} n_k^{-2 \nu_j/\nu_k} =\hbox{constant}.
\end{eqnarray}

\section{The \boldmath $m$-ion cases:\ $m = 1,\cdots, 4$}
\label{One_ion}

\subsection{\boldmath $m=1$}

In this single ion case,

\begin{eqnarray}
& &
n_1=\frac{p^2}{2} - (C x + K),\quad
p=-\frac{2 w'}{\nu_1 w},
\end{eqnarray}
where  $w$ satisfies the  Airy-type equation \cite{Fan},
\begin{eqnarray}
w'' - \frac{\nu_1^2}{2} ({C x +K}) w=0.
\end{eqnarray}

\subsection{\boldmath $m=2$}

Here, the $n_i,\,i =1,2$ are given by

\begin{eqnarray}
& &
n_1=\frac{p' - \nu_1 \left(\displaystyle\frac{p^2}{2} - C x -K\right)}{\nu_2-\nu_1},\quad
n_2=\frac{p' - \nu_2 \left(\displaystyle\frac{p^2}{2} - C x -K\right)}{\nu_1-\nu_2},\
\end{eqnarray}
and the second order ODE for $p(x)$ has degree one in $p''$.
All such second order ordinary differential equations
which have the Painlev\'e property have been
classified  by
Painlev\'e \cite{PaiBSMF} and Gambier  \cite{15}.

The first case
\begin{eqnarray}
& &
\nu_1:\nu_2=1:-2,\quad c_1=0,\quad c_2=0,\quad
p'' + \nu_1 p p' - \nu_1^2 p^3 + 2 K \nu_1^2 p=0
\end{eqnarray}
belongs to Class 10 of Gambier and its general solution is elliptic with
\begin{eqnarray}\fl
& &
p=-2 \frac{w'}{\nu_2 w},\quad
n_1=k_1 w,\quad
n_2=k_2 w^{-2},\quad
\frac{{w'}^2}{2 \nu_1^2 w^2} -k_1 w - k_2 w^{-2} - K=0.
\end{eqnarray}

The second case
\begin{eqnarray}
& &
\nu_1:\nu_2=1:-1,\quad
p''-\frac{1}{2} \nu_1^2 p^3 +\nu_1^2 (C x +K) +\nu_1 (c_1-c_2)=0
\end{eqnarray}
may be reduced to the  Painlev\'e II equation
\begin{eqnarray}
& &
\frac{\D^2 U}{\D X^2} = 2 U^3 + X U + \alpha
\label{eqPII}
\end{eqnarray}
under  the scaling transformation
\begin{eqnarray}
& &
p=\frac{2 k}{\nu_1}U,\quad
x=\frac{X}{k}-\frac{K}{C},\quad
k^3=- \nu_1^2 C,\quad
2 \alpha=\frac{c_1-c_2}{C}.
\label{eqm=2PII}
\end{eqnarray}

The third case
\begin{eqnarray}
& &
\nu_1:\nu_2=1:2,\quad
p''- 3 \nu_1 p p' + \nu_1^2 p^3 -2 \nu_1^2 (C x +K)
 +\nu_1 (c_1 + 2 c_2)=0
\end{eqnarray}
belongs to Class 5 of Gambier  and is linearizable with
\begin{eqnarray}
& &
p=-\frac{2 w'}{\nu_2 w},\quad
w''' -2 \nu_1^2 (C x +K) w' - \nu_1 (c_1 + 2 c_2)w=0.
\end{eqnarray}

\subsection{\boldmath $m =3$}


The three cases to be considered are:
\begin{eqnarray}
& &
\left.
\begin{array}{ll}
\displaystyle
\hbox{(3a): }
\quad \nu_1:\nu_2:\nu_3 = 1:-2:-1,\quad &(c_1,c_2,c_3)=(0,0,1) C,
\\[1ex]
 \displaystyle
\hbox{(3b): }
\quad \nu_1:\nu_2:\nu_3 = 1:-2:-1,\quad &(c_1,c_2,c_3)=(0,3,-2) C,
\\[1ex]
 \displaystyle
\hbox{(3c): }
\quad \nu_1:\nu_2:\nu_3 = 1:-2:2,\quad &(c_1,c_2,c_3)=(0,0,1) C.
\end{array}
\right\rbrace
\end{eqnarray}

The  class of equations to which
the ODE (\ref{eqode_p_orders1234})$_3$  belongs,  namely
\begin{eqnarray}
& &
-p^2 p''' + b p p' p'' + d {p'}^3 + h p^3 p''
 + k p^2 {p'}^2 + q p^4 p' + f p^6\\[2ex]
 &&
\qquad\qquad\qquad\qquad\quad\qquad\qquad\qquad\qquad\,\,
\mbox{}+ \hbox{subdominant terms}=0,\nonumber
\end{eqnarray}
has been investigated  in
\cite{ChazyThese}-\cite{KessiAdjabiOrder3Deg1},
but the results therein are insufficient to cover the above three cases.
For $C=K=0$,
the integration by elliptic functions (see (\ref{eqBriotBouquetW}))
was known to Chazy \cite{ChazyThese}.
For arbitrary values of $C$ and $K$,
we proceed directly with the original system (\ref{eqSystem_ni_p}).

In the first case, two $c_i$ vanish,
so that  $p$ obeys a second order ODE
and a birational transformation exists between $p$ and $w^2$ with
\begin{eqnarray}
& &
\hbox{(3a): }
\left\lbrace
\begin{array}{ll}
\displaystyle{
p=-2\frac{w'}{\nu_2 w},\quad
n_1=k_1 w,\quad
n_2=k_2 w^{-2},
} \\[2ex]
 \displaystyle{
w''-\frac{{w'}^2}{2 w} + \nu_1^2
  \left(-2 k_1 w^2 - (C x +K) w +\frac{k_2}{w} \right)=0,
} \\[2ex]
 \displaystyle{
3 k_2 \nu_1^2 w^{-2}=
\frac{p'' -\nu_1^2p^3/2 + \nu_1^2 (C x +K) p -\nu_1 C}{p}.
}
\end{array}
\right.
\end{eqnarray}
The ODE for $w$ belongs to Class 34 of Gambier
and its general solution is an algebraic transform of the PII
function $U$  (\ref{eqPII}) with
\begin{eqnarray}
& &
\hbox{(3a): }
\left\lbrace
\begin{array}{ll}
\displaystyle{
k_1 \nu_1^2 w= \varepsilon k \frac{\D U}{\D x} + k^2 U^2
 -\frac{\nu_1^2}{2}(C x +K),\quad
x=\frac{X}{k}-\frac{K}{C},\quad
} \\[2ex]
 \displaystyle{
k^3=- \nu_1^2 C,\quad
8 k_1^2 k_2= \frac{(2 \alpha+\varepsilon)^2 C^2}{\nu_1^2},\quad
\varepsilon^2=1.
}
\end{array}
\right.
\label{eq3asol}
\end{eqnarray}

In the second case,
since only one of the $c_i$ vanishes,
the ODE for $w$ is of third order
and a birational transformation links $p$ and $w$ with
\begin{eqnarray}
& &
\hbox{(3b): }
\left\lbrace
\begin{array}{ll}
\displaystyle{
p=-2\frac{w'}{\nu_2 w},\quad
n_1=k_1 w,
} \\[3ex]
\displaystyle
w''' + \nu_1^2 \left[- 6 k_1 w w' - 2 (C x +K) w' -4 C w\right]=0,
 \\[2ex]
 \displaystyle{
6 k_1 \nu_1^2 w =
\frac
{p'' + 3 \nu_1 p p' + \nu_1^2 p^3 -2 \nu_1^2 (C x + K) p -4 \nu_1 C}{p}.
}
\end{array}
\right.
\end{eqnarray}
The ODE for $w$ belongs to Class XIII of Chazy \cite{ChazyThese}
and  admits the first integral
\begin{eqnarray}\fl
& &
\hbox{(3b): }
\left\lbrace
\begin{array}{ll}
\displaystyle{
K_2=\left(w + (C x +K) k_1^{-1}\right) w'' -\frac{{w'}^2}{2}
} \\ \displaystyle{
\phantom{1234567}
  - \frac{C w'}{k_1} -2 k_1 \nu_1^2 w^3
  - 4 \nu_1^2 (C x + K) w^2
  - 2 \nu_1^2 (C x + K)^2 \frac{w}{k_1},
} \\[2ex]
 \displaystyle{
\phantom{123}
= \nu_1 k_1^{-2}
n_1 \left[\nu_1 (C x + K) n_3 - C p - \nu_1 n_1 n_2 \right].
}
\end{array}
\right.
\end{eqnarray}
This second order ODE for $w$
again belongs to Class 34 of Gambier and it may be
integrated in terms of the PII function $U$ with
\begin{eqnarray}
& &
\hbox{(3b): }
\left\lbrace
\begin{array}{ll}
\displaystyle
k_1 \nu_1^2 w=\varepsilon k \frac{\D U}{\D x} + k^2 U^2,\\[3ex]
\displaystyle
x=\frac{X}{k}-\frac{K}{C},\quad
k^3= 2 \nu_1^2 C,\quad
k_1^2 K_2 = - 2 C^2 \alpha(\alpha + \varepsilon),\quad\varepsilon^2=1.
\end{array}\right.
\end{eqnarray}
The third case, for which again two  $c_i$ vanish,
follows the same pattern as the first case and we have
\begin{eqnarray}
& &
\hbox{(3c): }
\left\lbrace
\begin{array}{ll}
\displaystyle{
p=-2\frac{w'}{\nu_3 w},\quad
n_1=k_1 w^{-1},\quad
n_2=k_2 w^2,\quad
} \\[2ex]
 \displaystyle{
w'' + \nu_1^2 \left[- 4 k_2 w^3 -2 (C x+K) w - k_1\right]=0,
} \\[2ex]
 \displaystyle{
12 k_2 \nu_1^2 w^2=\frac
{p'' - 3 \nu_1 p p' + \nu_1^2 p^3 -2 \nu_1^2 (C x+K) p + 2 \nu_1 C }{p},
}
\end{array}
\right.
\label{eq3csol}
\end{eqnarray}
so that a PII  equation is obtained for $w$.

It is remarked that, in the three cases considered above,
the first integral of the third order ODE
(\ref{eqode_p_orders1234})$_3$ for $p$
is a polynomial of third degree in $p''$.
This underlines the
advantage in proceeding with $w$ rather than $p$ in our analysis.

\subsection{\boldmath $m = 4$}

In the unique case to be integrated,
\begin{eqnarray}
& &
\nu_1:\nu_2:\nu_3:\nu_4 = 1:-2:-1:2,\quad
c_1=c_2=c_3=c_4=0,
\end{eqnarray}
the four first integrals immediately yield the solution,
\begin{eqnarray}
& &
\left\lbrace
\begin{array}{ll}
\displaystyle{
p=-\frac{w'}{2 \nu_2 w},\quad
n_1=k_1 w,\quad
n_2=k_2 w^{-2},\quad
n_3=k_3 w^{-1},\quad
n_4=k_4 w^2,
} \\[2ex]
 \displaystyle{
\frac{{w'}^2}{2 \nu_1^2 w^2}
- k_1 w - k_2 w^{-2} - k_3 w^{-1} - k_4 w^2 - K=0  ,
}
\end{array}
\right.
\end{eqnarray}
so that $w$ in general is an elliptic function.

The case when $m$ is arbitrary and all the $c_i$'s vanish
is similarly reduced to a first order ODE integrable by a quadrature,
however with a multi-valued solution when $m>4$.

\section{Conclusion}

The analysis presented here has been used to isolate, in particular,
underlying Painlev\'e II structure in the 3-ion electrodiffusion model.
This has potential application to the construction of upper and lower
solutions to two-point boundary value problems \cite{19}. The integrable
nature of the model in these cases also allows the application of B\"acklund
transformations \cite{20}.

\section*{Acknowledgement}
Robert Conte warmly thanks the Australian Research Council Centre
of Excellence for Mathematics and Statistics of Complex Systems
at the University of New South Wales for support for the visit
in which this work was started.

\section*{Appendix. Conditions for the absence of movable logarithms}

After the coefficient $p_0$ of the series (\ref{eqpLaurent})
has been computed,
the recurrence relation for subsequent $p_k$, namely
\begin{equation}
\forall k\ge 1:\
E_k \equiv P(k) p_k + Q_k(\{p_l\ \vert \ l<k \}) = 0,
\label{eqMethodPole5}
\end{equation}
depends linearly on $p_k$
and nonlinearly on the previously computed coefficients $p_l$.
Whenever the positive integer $k$ is a Fuchs index $r$ of the
$m$-th order ODE (\ref{eqode_p_orderm}),
the coefficient $P(r)$ vanishes
and the condition $Q_r=0$ must be enforced
in order to avoid movable logarithms.
For the cases isolated in (\ref{eqlistmnu})
and for each of the $m$ families $p_0=-2/\nu_j$,
these conditions $Q_r=0$ deliver the following.\\

$m=2$:
\begin{eqnarray}
& &
\left\lbrace
\begin{array}{ll}
\displaystyle{
\nu_1:\nu_2=1:-2,\ j=1,\ Q_6=c_2 (c_1+2 c_2),
}
\\
\displaystyle{
\nu_1:\nu_2=1:-2,\ j=2,\ Q_3=c_1,
}
\\
\displaystyle{
\nu_1:\nu_2=1:-1,\ j=1,\ Q_4=0,
}
\\
\displaystyle{
\nu_1:\nu_2=1:-1,\ j=2,\ Q_4=0,
}
\\
\displaystyle{
\nu_1:\nu_2=1: 2,\ j=1,\ \hbox{no condition},
}
\\
\displaystyle{
\nu_1:\nu_2=1: 2,\ j=2,\ Q_1=0.
}
\end{array}
\right.
\end{eqnarray}

$m=3$:
\begin{eqnarray}
& &
\left\lbrace
\begin{array}{ll}
\displaystyle{
\nu_1:\nu_2:\nu_3= 1:-2:-1,\ j=1,\ Q_4=0,\ Q_6=c_2(c_1+2 c_2+3 c_3),
}
\\
\displaystyle{
\nu_1:\nu_2:\nu_3= 1:-2:-1,\ j=2,\ Q_1=0,\ Q_3=c_1,
}
\\
\displaystyle{
\nu_1:\nu_2:\nu_3= 1:-2:-1,\ j=3,\ Q_4=0,
}
\\
\displaystyle{
\nu_1:\nu_2:\nu_3= 1:-2: 2,\ j=1,\ Q_6=c_2(5 c_1 + 10 c_2 + 6 c_3),
}
\\
\displaystyle{
\nu_1:\nu_2:\nu_3= 1:-2: 2,\ j=2,\ Q_3=c_1,\ Q_4=0,
}
\\
\displaystyle{
\nu_1:\nu_2:\nu_3= 1:-2: 2,\ j=3,\ Q_1=0,\ Q_4=c_2 p_1.
}
\end{array}
\right.
\end{eqnarray}

$m=4$: $\nu_1:\nu_2:\nu_3:\nu_4 = 1:-2:-1:2$,
\begin{eqnarray}
& &
\left\lbrace
\begin{array}{ll}
\displaystyle{
j=1,\ Q_4=0,\ Q_6=c_2(5 c_1+10 c_2+15 c_3+6 c_4),
}
\\
\displaystyle{
j=2,\ Q_1=0,\ Q_3=c_1,\ Q_4=(3 c_1+2 c_4) p_1,
}
\\
\displaystyle{
j=3,\ Q_4=0,\ Q_6=c_4(5 c_3+10 c_4+15 c_1+6 c_2),
}
\\
\displaystyle{
j=4,\ Q_1=0,\ Q_3=c_3,\ Q_4=(3 c_3+2 c_2) p_1.
}
\end{array}
\right.
\end{eqnarray}
Since the coefficient $p_1$ must remain arbitrary,
this leads to the list (\ref{eqPositiveLog}).


\vfill\eject
\end{document}